\documentclass[11pt,a4paper]{article}

\usepackage{amsmath}
\usepackage{amsfonts}
\usepackage{amssymb}
\usepackage{graphicx,psfrag,color}
\usepackage{subfigure}
\usepackage{hyperref}

\usepackage{enumerate}

\usepackage[left=2cm,top=2cm,right=2cm,bottom=2cm]{geometry}

\begin{document}
\noindent \textcolor{blue}{\textbf{Phys. Rev. D \textbf{92} (2015) 103512}}
\begin{center}
\Large{\textbf{Consistency of the nonflat $\Lambda$CDM model with the new result from BOSS}} \\[0.5cm]
 
\large{\textbf{ Suresh Kumar\footnote{ suresh.kumar@pilani.bits-pilani.ac.in}}}
\\[0.5cm]

\textit{ Department of Mathematics, BITS-Pilani, Pilani Campus, Rajasthan-333031, India.}

\end{center}

\vspace{.2cm}

\noindent \textbf{\large{Abstract:}}  Using 137,562 quasars in the redshift range $2.1\leq z\leq3.5$ from the Data Release 11 (DR11) of the Baryon Oscillation Spectroscopic Survey (BOSS) of Sloan Digital Sky Survey (SDSS)-III, the BOSS-SDSS collaboration estimated  the expansion rate $H(z=2.34)=222\pm7$ km/s/Mpc of Universe, and reported that this value is in tension with the predictions of flat $\Lambda$CDM model at around 2.5$\sigma$ level. In this paper, we briefly describe some attempts made in the literature to relieve the tension, and show that the tension can naturally be alleviated in nonflat $\Lambda$CDM model with positive curvature. We also perform the observational consistency check by considering the constraints on the nonflat $\Lambda$CDM model from Planck,WP and BAO data. We find that the nonflat $\Lambda$CDM model constrained with Planck+WP data fits better to the line of sight measurement $H(z=2.34)=222\pm7$ km/s/Mpc, but only at the expense of still having a poor fit to the BAO transverse measurements.


\section{Introduction}
The overall energy budget of the Universe is dictated by dark matter and dark energy with a minor contamination from baryonic matter, where the dark energy is assumed to dominate the cosmic landscape causing the acceleration of Universe. In the last two decades, though we have witnessed high precision data/observations in cosmology, we are still struggling to find the suitable candidates for dark matter and dark energy from fundamental physics. The standard flat $\Lambda$CDM (cosmological constant + cold dark matter) model, however,  has been remarkably successful in describing cosmic acceleration in  the present Universe by fitting the observational data very well \cite{PlanckXXVI}. In this model, the cosmological constant ($\Lambda$) mimics the dark energy characterized by the equation of state parameter  $w_\Lambda=p_\Lambda/\rho_\Lambda=-1$, where $p_\Lambda$ and $\rho_\Lambda$ are respectively the pressure and energy density of vacuum. This model, however, suffers from several theoretical problems \cite{peebles03,copeland06}. This paves the way for studying alternative models of dark energy. 

\begin{figure}[htb!]\centering
\includegraphics[width=14cm]{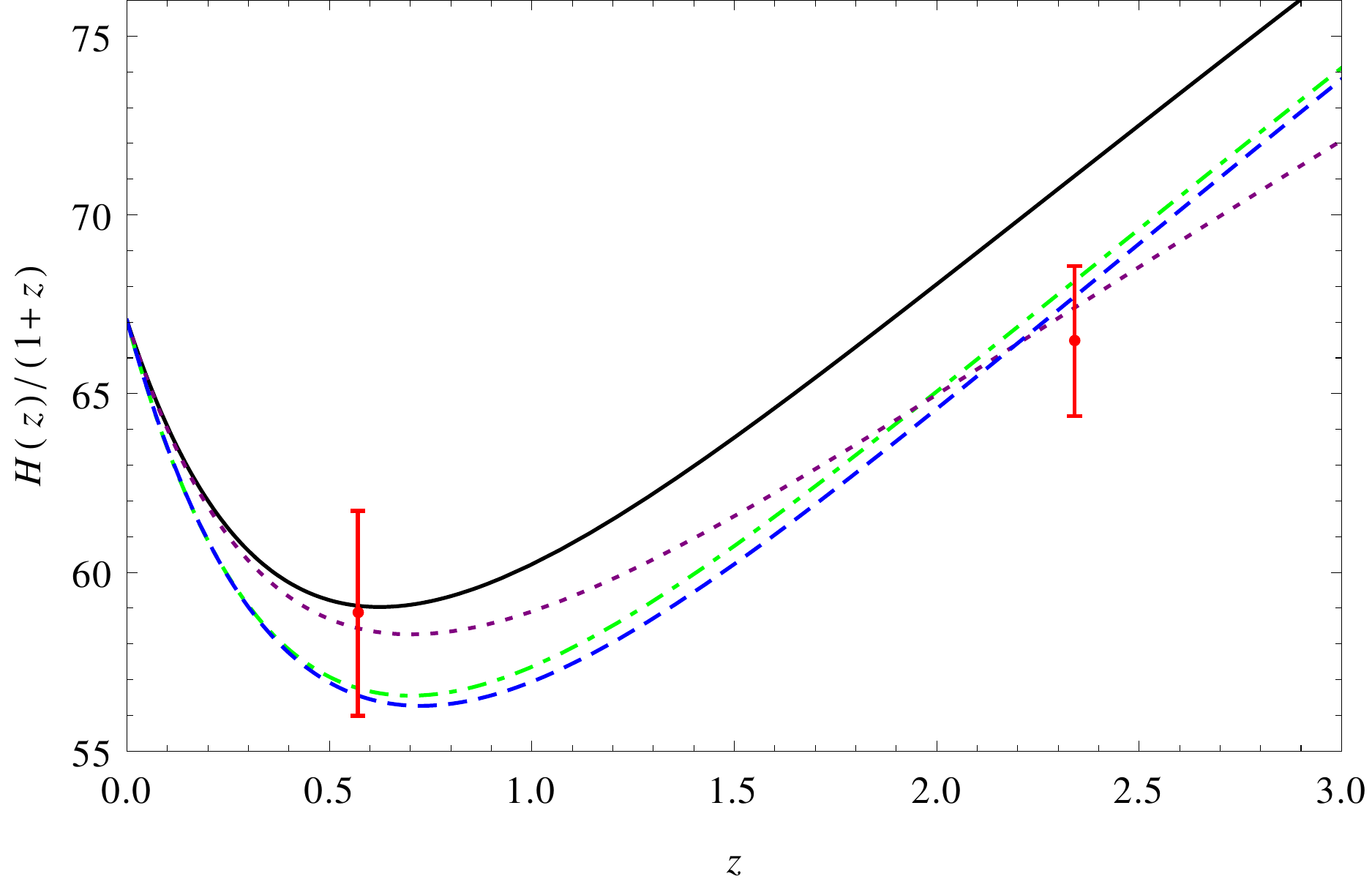}
\caption{\footnotesize{Plots of $H(z)/(1+z)$ curves for various models are shown in the redshift range $0\leq z\leq 3$. Solid Black curve: Flat $\Lambda$CDM model \eqref{eq1}.  Dotted Purple curve: Interacting dark sector model \eqref{int1}. Dashed Blue curve: Braneworld model \eqref{eq5}. Dotdashed Green curve: Nonflat $\Lambda$CDM model \eqref{eq6}. The error bars in Red color represent $H(z=0.57)=92.4\pm4.5$km/s/Mpc (left) \cite{samushia13} and $H(z=2.34)=222\pm 7$km/s/Mpc (right) \cite{delubac15}, both scaled by the factor $1/(1+z)$.}}
\label{fig:1}
\end{figure}

Model independent measurements of Hubble parameter $H(z)$ at different redshifts provide a useful and counter test for other observational probes in concordance cosmology. For instance, Verde et al. \cite{verde14}  compared cosmology independent measurements of $H(z)$ in the redshift range $0 . 1 \lesssim z < 1 . 2$ with the Cosmic Microwave Background (CMB)-derived expansion history predictions, and deduced that there is no tension between Planck and cosmology independent measurements of the $H ( z )$ for the flat $\Lambda$CDM model. In a recent study, Delubac et al.\cite{delubac15} estimated  $H(z=2.34)=222\pm7$ km/s/Mpc by using 137,562 quasars in the redshift range $2.1\leq z\leq3.5$ from the Data Release 11 (DR11) of the Baryon Oscillation Spectroscopic Survey (BOSS) of SDSS-III, and reported that this value is in tension with Planck+WP constraints on flat $\Lambda$CDM model at around 2.5$\sigma$ level. In fact, this value of $H(z=2.34)$ is lower than the value predicted by the flat $\Lambda$CDM model. For precise details of the inconsistency, the readers are advised to see the Ref. \cite{delubac15}. Here, just for the sake of a first up handwaving  demonstration of the inconsistency\footnote{We shall consider proper fitting of the flat $\Lambda$CDM model with the Planck+WP data in Section \ref{4}, and illustrate the said inconsistency there.}, we choose the best fit estimates $H_0=67.04$ km/s/Mpc and $\Omega_{\rm m0}=0.3183$ for the present values of Hubble parameter $H$ and matter density parameter $\Omega_{\rm m}$ in the flat $\Lambda$CDM model, given by Planck+WP data\cite{PlanckXXVI}.  We plot the expansion history 
\begin{equation}\label{eq1}
H(z)=H_{0}\sqrt{\Omega_{\rm m0}\left(1+z\right)^{3}+\Omega_{\Lambda}},
\end{equation}
of the flat $\Lambda$CDM model (solid Black curve) in Figure \ref{fig:1} scaled by the factor $1/(1+z)$, where the error bar on left represents $H(z=0.57)=92.4\pm4.5$ km/s/Mpc \cite{samushia13} while the right one stands for $H(z=2.34)=222\pm7$ km/s/Mpc \cite{delubac15}, both scaled by the factor $1/(1+z)$. The other curves in Figure \ref{fig:1} correspond to other models, and will be discussed  in the subsequent sections. We see that the $H(z)/(1+z)$ curve  (solid Black) of flat $\Lambda$CDM model passes through the left error bar but fails to pass through the right one.  Therefore, the framework of the flat $\Lambda$CDM cosmology is not adequate to be consistent with the result from BOSS. This in turn implies the need to explore deviations from the flat $\Lambda$CDM cosmology. One such natural deviation is the nonflat $\Lambda$CDM model. The purpose of this study is to demonstrate that this natural extension of flat $\Lambda$CDM model is capable of explaining the lower value of Hubble parameter at higher redshift as reported in BOSS measurements. In the next section, we briefly describe some attempts/suggestions made in the literature to justify the lower value of Hubble parameter at higher redshift from BOSS while we present, in Section \ref{3}, the motivation for studying the nonflat $\Lambda$CDM model. In Section \ref{4}, we discuss the observational constraints on the nonflat $\Lambda$CDM model from some recent observational data sets, and do its observational consistency check with the BOSS measurements while discussing its related strengths and weaknesses in detail. In Section \ref{5}, we present concluding remarks of the study.

\section{Attempts to relieve the tension}\label{2}

\subsection{Interacting dark sector models}\label{2.1}
Following \cite{he09,costa14}, Abdalla et al.\cite{abdalla14} considered some models for the dark sector interaction, and showed that a simple phenomenological interaction in the dark sector provides a good explanation for the deviation from the flat $\Lambda$CDM cosmology, naturally accommodating the Hubble parameter obtained by BOSS. One such model reads as
\begin{equation}
\label{int1}
H(z)=H_{0}\sqrt{\Omega_{\rm b0}\left(1+z\right)^{3}+\Omega_{\rm dm0}(1+z)^{3(1-\xi)}+(1+z)^{3(1+w)}\left(\Omega_{\rm de0}+\frac{\Omega_{\rm dm0}\xi}{\xi+w}\right)-\frac{\Omega_{\rm dm0}\xi}{\xi+w}(1+z)^{3(1-\xi)}},
\end{equation}
where $\Omega_{\rm b0}$, $\Omega_{\rm dm0}$ and $\Omega_{\rm de0}$ are the density parameters of baryonic matter, dark matter and dark energy, respectively at $z=0$. Further, $\xi$ is the coupling constant coming from interaction term, and $w$ is equation of state parameter of dark energy (See \cite{he09,abdalla14} for more details). Here, for a quick handwaving illustration, we plot the expansion history \eqref{int1} scaled by $1/(1+z)$ in Figure \ref{fig:1} (dotted Purple curve) for selected values of the constants given by $H_0=67.04$ km/s/Mpc, $\Omega_{\rm b0}=0.048$, $\Omega_{\rm dm0}=0.2703$, $\xi = 0.05$ and $w = -1.01$. We see that $H(z)/(1+z)$ curve (dotted Purple) for the dark sector interaction model \eqref{int1} passes through both the  error bars. So the lower value of $H(z)$ at higher redshift from BOSS could be an indication of the interaction in the dark sector as concluded in \cite{abdalla14}.

\subsection{Models of screened dark energy}\label{2.2}
Sahni et al. \cite{sahni14} pointed out that the new result from BOSS could be an indication of the existence of dark energy with negative energy density at higher redshifts. Also, they  mentioned that the general relativity framework of flat $\Lambda$CDM model may not be adequate, and suggested that the lower values of $H(z)$ at higher redshifts can be achieved by considering a dynamically evolving term with minus sign in the expansion history \eqref{eq1} of the flat $\Lambda$CDM model, that is,
\begin{equation}
\label{eqn:Hzscreened}
H(z)=H_{0}\sqrt{\Omega_{\rm m0}\left(1+z\right)^{3}+\Omega_{\Lambda}-f(z)}.
\end{equation}
Here $f(z)>0$ is the dynamically evolving counter-term, which evolves in such a way that it screens or compensates the cosmological constant in the past, and thereby gives rise to lower values of $H(z)$ at higher redshifts. Moreover, $f(z)$ must grow monotonically with $z$ but at a rate slower than $(1+z)^3$ in order to preserve the matter-dominated regime. To illustrate the screening mechanism, they considered a Braneworld model proposed in \cite{sahni03}, given by the equations
\begin{equation}\label{eq5}
H(z)=H_{0}\sqrt{\Omega_{\rm m0}\left(1+z\right)^{3}+\Omega_{\Lambda}-\left[2\sqrt{\Omega_{l}}\sqrt{\Omega_{\rm m0}\left(1+z\right)^{3}+\Omega_{\Lambda}+\Omega_{l}}-2\Omega_{l}\right]},
\end{equation}
\[\Omega_{\Lambda}=1-\Omega_{m0}+2\sqrt{\Omega_{l}},\]
where $\Omega_{\rm m0}$, $\Omega_{\Lambda}$ and $\Omega_{l}$ are the density parameters at $z=0$ as per the Braneworld scenario considered in \cite{sahni03}. The plot of $H(z)/(1+z)$ for the Braneworld model \eqref{eq5} is shown in Figure \ref{fig:1} by dashed Blue curve for selected values of the parameters given by $H_0=67.04$ km/s/Mpc, $\Omega_{\rm m0}=0.3183$ and $\Omega_{l}=0.06$. Again, we see that $H(z)/(1+z)$ curve (dashed Blue) for the Braneworld model \eqref{eq5} passes through both the  error bars. Therefore, the introduction of screening term in the evolution history of flat $\Lambda$CDM  model is a viable option to explain the lower value of $H(z)$ at higher redshift from BOSS. 
\section{Nonflat $\Lambda$CDM model with positive curvature}\label{3}
It is evident that one can use numerous forms of the screening term $f(z)$ in equation \eqref{eq5} to obtain a desired history. Every time, one needs to explain the physics of the screening term. Among the numerous possibilities, we point out that the curvature term $\Omega_{k}(1+z)^2$ with $k>0$\footnote{In standard notations, $\Omega_k<0$ for $k>0$.} is a natural and plausible choice for $-f(z)$, where $\Omega_{k}$ is the curvature density parameter at $z=0$.  In other words, we advocate the consideration of the nonflat $\Lambda$CDM model with positive curvature, given by
\begin{equation}
\label{eq6}
H(z)=H_{0}\sqrt{\Omega_{\rm m0}\left(1+z\right)^{3}+\Omega_{\Lambda}+\Omega_{k}(1+z)^2},
\end{equation}
for dealing with the new result from BOSS. Notice that this model is within the framework of general relativity, and it does not invoke anything hypothetical or fancy like dark energy with negative energy density.  

In view of \eqref{eq6}, one may interpret that the curvature term $\Omega_{k}(1+z)^2$ with $k>0$ screens the dark energy given by cosmological constant, and balances the contribution of cosmological constant at the redshift $z=(\Omega_{\Lambda}/|\Omega_{k}|)^{1/2}-1$. Therefore,  the curvature term is responsible for giving rise to lower values of $H(z)$ at higher redshifts in the nonflat $\Lambda$CDM model. Also, $\Omega_{k}(1+z)^2$ evolves at a rate slower than the term $\Omega_{\rm m0}\left(1+z\right)^{3}$, and thereby preserves the matter regime. So the curvature term $\Omega_{k}(1+z)^2$ with $k>0$ fits well in the screening mechanism suggested by Sahni et al.\cite{sahni14}. However, one does not need this kind of interpretation/explanation in case of the nonflat $\Lambda$CDM model \eqref{eq6} because the curvature term $\Omega_{k}(1+z)^2$ with $k>0$ has a physical meaning, and it does not require to be interpreted as a dark energy source with negative energy density. Now, we plot the expansion history \eqref{eq6} (scaled by $1/(1+z)$) in Figure \ref{fig:1} by dotdashed Green curve by choosing $H_0=67.04$ km/s/Mpc, $\Omega_{\rm m0}=0.3183$ and $\Omega_{k}=-0.1$. We notice that the $H(z)/(1+z)$ curve (dotdashed Green)  of the nonflat $\Lambda$CDM model with positive curvature passes through both the  error bars. Therefore, the nonflat $\Lambda$CDM model with positive curvature is capable of explaining the lower value of $H(z=2.34)$ suggested by BOSS.

The above idea of considering nonflat $\Lambda$CDM model with positive curvature seems simple and natural to satisfy the result from BOSS as it works well within the framework of general relativity, and does not require any modification of general relativity. Considering our current knowledge of fundamental physics, it is more appealing than the idea of invoking dark energy with negative energy density, or the phenomenological proposals such as the interaction of dark matter and dark energy as described in the previous section.  Moreover, the dark matter and dark energy interaction model with additional free parameters oversteps the bounds of Occam's rajor whereas curvature provides a simpler solution. However, curvature models face a direct confrontation with the inflation paradigm which predicts almost a spatially flat Universe \cite{PlanckXXVI,aub14}.  We shall comment more on this issue later in Section \ref{5}. Next, it should be noted that we have checked the consistency of the nonflat $\Lambda$CDM model with the new result from BOSS by choosing $\Omega_{k}=-0.1$ at our own just for a handwaving illustration. So the nonflat $\Lambda$CDM model with positive curvature, of course, requires to pass the filter of observational data from different probes before being accepted as a reliable model to deal with the new result from BOSS. Therefore, in the next section, we discuss observational constraints on the nonflat $\Lambda$CDM model \eqref{eq6}, and do its observational consistency check with the new result from BOSS.

\section{Observational constraints }\label{4}
The measurements of CMB temperature power spectrum are known to suffer from ``geometrical degeneracy" \cite{bond97,zal97}. In fact, almost identical CMB temperature power spectra are observed in models with identical physical matter densities, primordial spectra  and angular diameter distance to the last scattering surface. This nearly perfect degeneracy is broken by using various approaches \cite{PlanckXXVI}. For instance, analyzing the integrated Sachs-Wolfe (ISW) effect on large angular scales and gravitational lensing of the
CMB spectrum \cite{stom99}, considering additional probes of
late time Universe such as measurements of BAO, SN Ia and $H_0$ \cite{sper07}, are to name a few. In Table \ref{tab:results}, we show observational constraints on $\Omega_k$ obtained in the literature in recent years by confronting the nonflat $\Lambda$CDM model to various observational data sets. The readers may refer to the respective references for the details of the data sets. It is interesting to observe that all the data sets mentioned in Table \ref{tab:results} predict negative mean values of $\Omega_k$. This observation is useful in the context of our study because the nonflat $\Lambda$CDM model with $\Omega_k<0$ may yield lower values of $H(z)$ at higher redshifts, as discussed in the previous section.

\begin{table}[h]
\caption{Observational constraints on $\Omega_k$.} 
\begin{center}
\begin{tabular}{|lll|}
\hline Data&$\quad \Omega_k$&Reference\\ \hline 
WMAP7+ACT	& $-0.031 \pm 0.026$ (68\% C.L.) & \cite{Sievers13}\\
WMAP7+SPT	& $-0.0014 \pm 0.0172$ (68\% C.L.) & \cite{van12}\\
WMAP7+SPT+K11	& $-0.0015 \pm 0.0146 $ (68\% C.L.) & \cite{van12}\\
Planck+WP	& $-0.037_{-0.049} ^{+0.043}$ (95\% C.L.) & \cite{PlanckXXVI} \\
Planck+WP+highL & $-0.042_{-0.048} ^{+0.043}$ (95\% C.L.) & \cite{PlanckXXVI}\\
Planck+WP+highL+BAO & $-0.0005_{-0.0066} ^{+0.0065}$ (95\% C.L.) &  \cite{PlanckXXVI}\\
Planck TT & $-0.052_{-0.055} ^{+0.049}$ (95\% C.L.) & \cite{Planck15}\\
Planck TT+lensing & $-0.005_{-0.017} ^{+0.016}$ (95\% C.L.) & \cite{Planck15}\\
Planck TT+lensing+BAO+JLA+$H_0$ & $-0.0001_{-0.0052} ^{+0.0054}$ (95\% C.L.) & \cite{Planck15}\\
Planck TT, TE, EE & $-0.040_{-0.041} ^{+0.038}$ (95\% C.L.) & \cite{Planck15}\\
Planck TT, TE, EE+lensing & $-0.004_{-0.015} ^{+0.015}$ (95\% C.L.) & \cite{Planck15}\\
\hline
\end{tabular}
\label{tab:results}
\end{center}
\end{table}
Since the BOSS collaboration has reported the tension with Planck+WP determined flat $\Lambda$CDM cosmology, it would be interesting to see how the Planck+WP data responds to the nonflat $\Lambda$CDM model. So we consider  the observational constraints on the parameters of the nonflat $\Lambda$CDM model \eqref{eq6} from the Planck+WP data, in addition to the joint Planck+WP+BAO data set. We utilize the publicly available chains provided by Planck 2013 collaboration \cite{PlanckXXVI} generated by using the Markov Chain Monte Carlo (MCMC) method coded in the publicly available package cosmoMC \cite{cos02}. 
\begin{table}[htb!]\small\centering
\caption{Constraints on parameter space of nonflat $\Lambda$CDM model \eqref{eq6} from Planck+WP and Planck+WP+BAO data. Mean values are shown with 68\% and 95\% C.L.}\label{tab:results1}
\begin{tabular}{|l|ll|ll|}
\hline 
Data $\rightarrow$ & \multicolumn{2}{|c|}{Planck+WP} &  \multicolumn{2}{c|}{Planck+WP+BAO}\\\hline
Parameters & Mean with errors & Bestfit& Mean with errors & Bestfit \\ \hline 
$\Omega_b h^2$ & $    0.02231_{-    0.00031-    0.00061}^{+    0.00060+    0.00060}$ & $    0.02249$&$0.02211^{+0.00030+0.00059}_{-0.00029-0.00057}$ & $    0.02202$
 \\[8pt]
$\Omega_c h^2$ & $0.1183^{+0.0027+0.0053}_{-0.0027-0.0052}$ & $    0.1181$&$0.1188^{+0.0027+0.0053}_{-0.0027-0.0052}$& $    0.1179$
 \\[8pt]
$100\theta_{MC}$ & $1.0416^{+0.0006+0.0013}_{-0.0006-0.0013}$ & $    1.0414$& $1.0414^{+0.0006+0.0013}_{-0.0006-0.0013}$ & $    1.0415$
 \\[8pt]
$\tau$ & $0.087^{+0.014+0.024}_{-0.012-0.026} $& $    0.094$& $0.091^{+0.013+0.027}_{-0.013-0.027}   $ & $    0.094$
 \\[8pt]
$n_s$ & $0.965^{+0.008+0.015}_{-0.007-0.015}   $& $    0.969$& $0.963^{+0.007+0.014}_{-0.007-0.015}$ & $    0.9632$
 \\[8pt]
${\rm{ln}}(10^{10} A_s)$ & $3.081^{+0.024+0.068}_{-0.027-0.047}$ & $    3.097$&  $3.089^{+0.025+0.052}_{-0.026-0.049}   $& $    3.094$
\\[8pt]
$\Omega_k$ & $-0.037^{+0.027+0.043}_{-0.017-0.048}  $ & $    -0.038$&$0.00003^{+0.00334+0.00658}_{-0.00330-0.00668}  $ & $    -0.00037$
 \\[8pt]
\hline

$\Omega_{\rm m0}$ & $0.457^{+0.069+0.181}_{-0.102-0.166}   $ & $    0.459$& $0.309^{+0.010+0.022}_{-0.011-0.020}   $ & $    0.305$
 \\[8pt]
$H_0$ & $56.3^{+5.1+11.0}_{-5.8-10.0}        $ & $   55.4$& $67.8^{+1.0+2.0}_{-1.0-2.1}        $ & $   67.8$
 \\[8pt]
${\rm{Age}}/{\rm{Gyr}}$ & $15.09^{+0.66+1.34}_{-0.67-1.26}     $ & $   15.16$& $13.80^{+0.13+0.27}_{-0.13-0.26}     $ & $   13.82$\\
\hline
\end{tabular}

\end{table}

\begin{center}
\begin{figure}[htb]
\includegraphics[width=17cm]{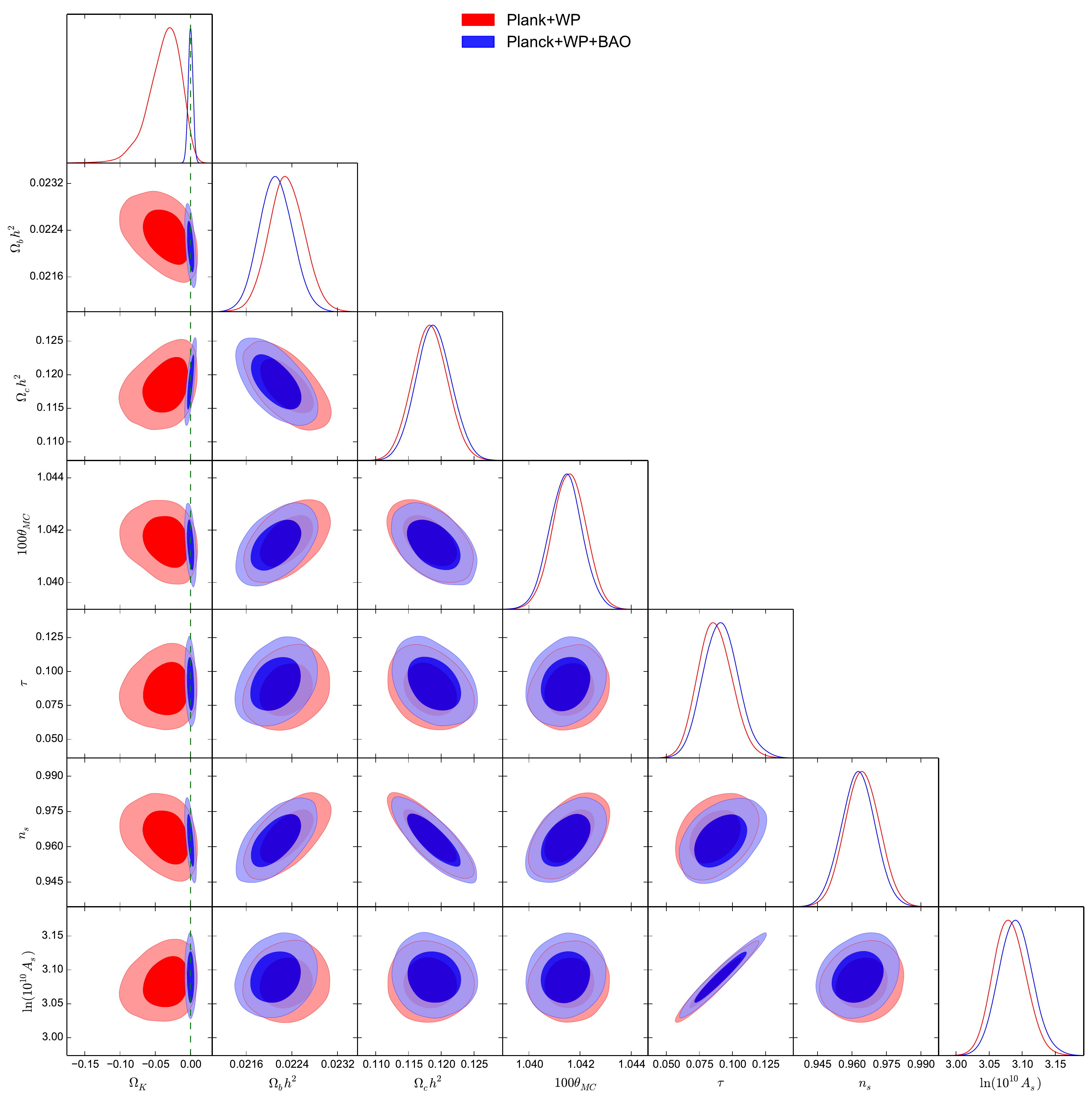}
\caption{The one-dimensional marginalized distribution, and two-dimensional contours  with $68\%$ C.L. and $95\%$ C.L. for the parameters of the nonflat $\Lambda$CDM model. The vertical dotted Green line stands for $\Omega_k=0$.}\label{fig:contour}
\end{figure}
\end{center}

To comply with the standards in the field, we explore the full parameter space of the nonflat $\Lambda$CDM model \eqref{eq6}. The constraints on its full parameter space 
$P\equiv\{\omega_{b},\omega_c, \Theta_{S},\tau, n_{s},\log[10^{10}A_{s}],\Omega_k\}$ and some derived parameters of interest
from the Planck+WP and Planck+WP+BAO data are displayed in Table \ref{tab:results1}.  In Figure \ref{fig:contour}, we show one-dimensional marginalized distribution and two-dimensional contours  with $68\%$ C.L. and $95\%$ C.L. for the parameters under consideration of the nonflat $\Lambda$CDM model. The vertical dotted Green line stands for $\Omega_k=0$.

From Table \ref{tab:results1} and Figure \ref{fig:contour}, we observe that Planck+WP+BAO data offer tight constraints on the parameter space of the nonflat $\Lambda$CDM model, in comparison to the constraints given by Planck+WP data. In Figure \ref{fig:contour}, we see that the major part of confidence regions of $\Omega_k$ corresponding to the Plank+WP data lie to the left of the vertical dotted Green line $\Omega_k=0$, which shows that the Plank+WP data favor the negative values of $\Omega_k$. Consequently, Planck+WP data yield negative mean and best fit values of $\Omega_k$, as may be seen in Table \ref{tab:results1}. On the other hand, Planck+WP+BAO data push the value of  $\Omega_k$ in the close vicinity of 0, and thus tend to favor the flat $\Lambda$CDM cosmology.

\begin{figure}[htb!]\centering
\includegraphics[width=8.3cm]{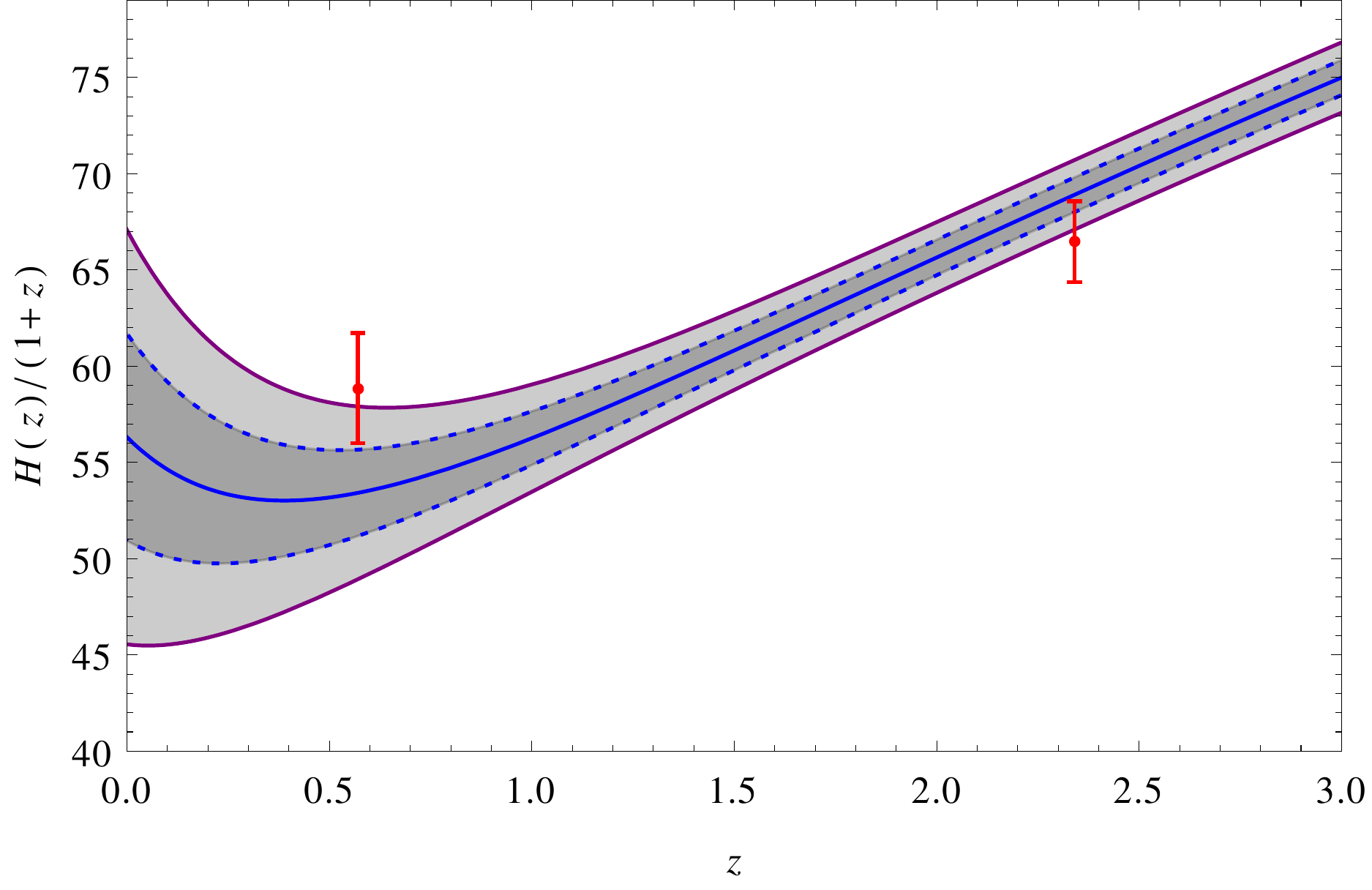}
\includegraphics[width=8.3cm]{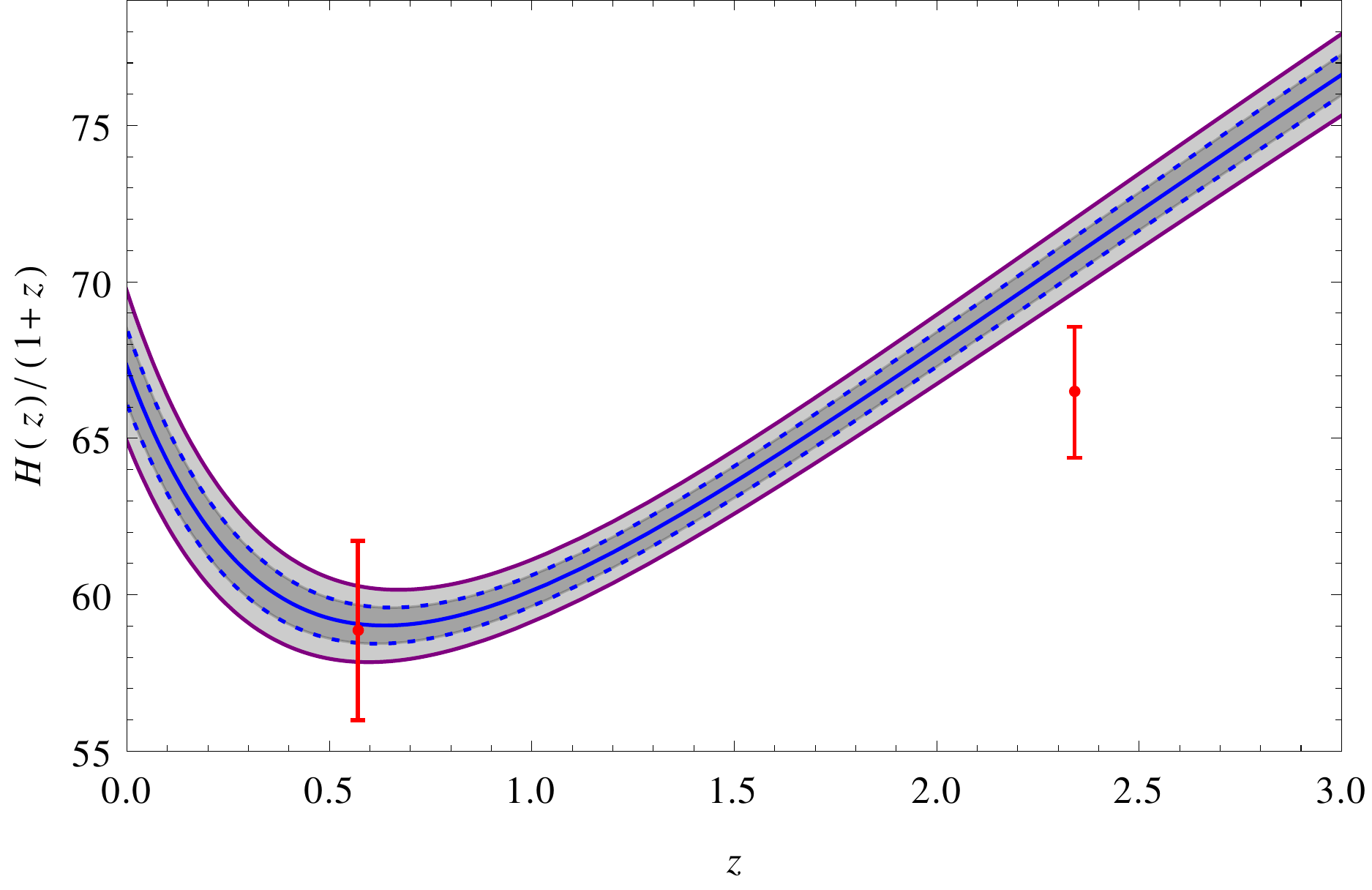}
\caption{\footnotesize{Plots of $H(z)/(1+z)$ curve with 1$\sigma$ and 2$\sigma$ error bands for nonflat (left panel) and flat (right panel) $\Lambda$CDM models constrained by Planck+WP data.}}
\label{fig:3}
\end{figure}

Now we turn our attention to the consistency check of the nonflat $\Lambda$CDM model \eqref{eq6} constrained from the Planck+WP data set with the new result from BOSS. In Figure \ref{fig:3}, we  plot $H(z)/(1+z)$ curve  with 1$\sigma$ and 2$\sigma$ error bands for nonflat (left panel) and flat (right panel) $\Lambda$CDM models constrained by Planck+WP data. We see that the nonflat $\Lambda$CDM model fits the new result from BOSS even at 1$\sigma$ level while the flat one fails to do so even at 2$\sigma$ level. Thus, the nonflat $\Lambda$CDM model with positive curvature is consistent with the new result from BOSS. However, it requires further testing with BAO transverse measurements. For, in BAO measurements, separations along the line of sight correspond to differences in redshift that depend on the Hubble parameter while separations transverse to the line of sight correspond to differences in angle that depend on the angular diameter distance, in addition to the radius of the sound horizon at the drag epoch when photons and baryons decouple. Since BAO also accurately measure the
transverse scale, it is required to have an observational consistency check of the nonflat $\Lambda$CDM model for the BAO transverse measurements too\footnote{The new result from BOSS or the lower Hubble parameter value measurement at higher redshift from BOSS as phrased earlier at many places in this study, is in fact a measurement along the line of sight. Hereafter, we shall refer it to as the line of sight high redshift BOSS measurement or simply the line of sight BOSS measurement while the measurements transverse to the line of sight shall be referred to as the BAO transverse measurements.}. Following \cite{aub14}, we utilize the transverse comoving distance $D_M(z)$ in connection to the BAO transverse measurements, which reads as
\begin{equation}
D_M(z)=\frac{c}{H_0}S_{k}\left(\frac{D_C(z)}{c/H_0}\right),
\end{equation}
where 
\begin{equation}
D_C(z)=\frac{c}{H_0}\int_{0}^{z}\frac{H_0}{H(z')}dz',
\end{equation}
is the line of sight distance, and
\begin{equation}
S_k(x)=\left\{ \begin{array}{lll}
         \sin(\sqrt{-\Omega_k}x)/\sqrt{-\Omega_k}\; ,& \mbox{ $\Omega_k<0$}\\
         \sinh(\sqrt{\Omega_k}x)/\sqrt{\Omega_k}\; ,& \mbox{ $\Omega_k>0$}\\
        x \;,& \mbox{ $\Omega_k=0$}.\end{array} \right.
\end{equation} 
For small values of $|\Omega_k|$, we have
\begin{equation}
D_M(z)\approx D_C(z)\left[1+\frac{1}{6}\Omega_k\left(\frac{D_C(z)}{c/H_0}\right)^2+\frac{1}{120}\Omega_k^2\left(\frac{D_C(z)}{c/H_0}\right)^4\right].
\end{equation}
\begin{figure}[htb!]\centering
\includegraphics[width=8.3cm]{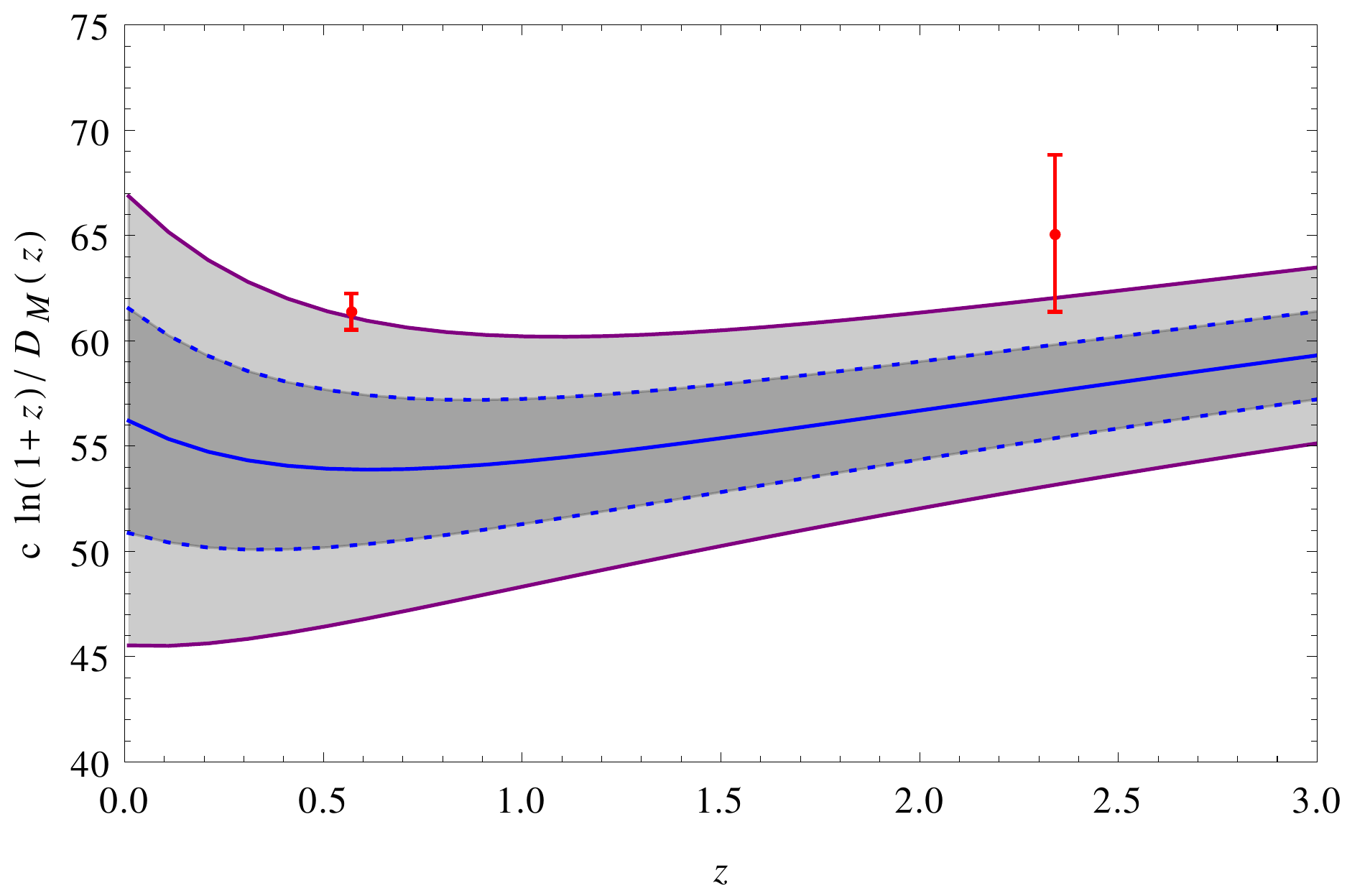}
\includegraphics[width=8.3cm]{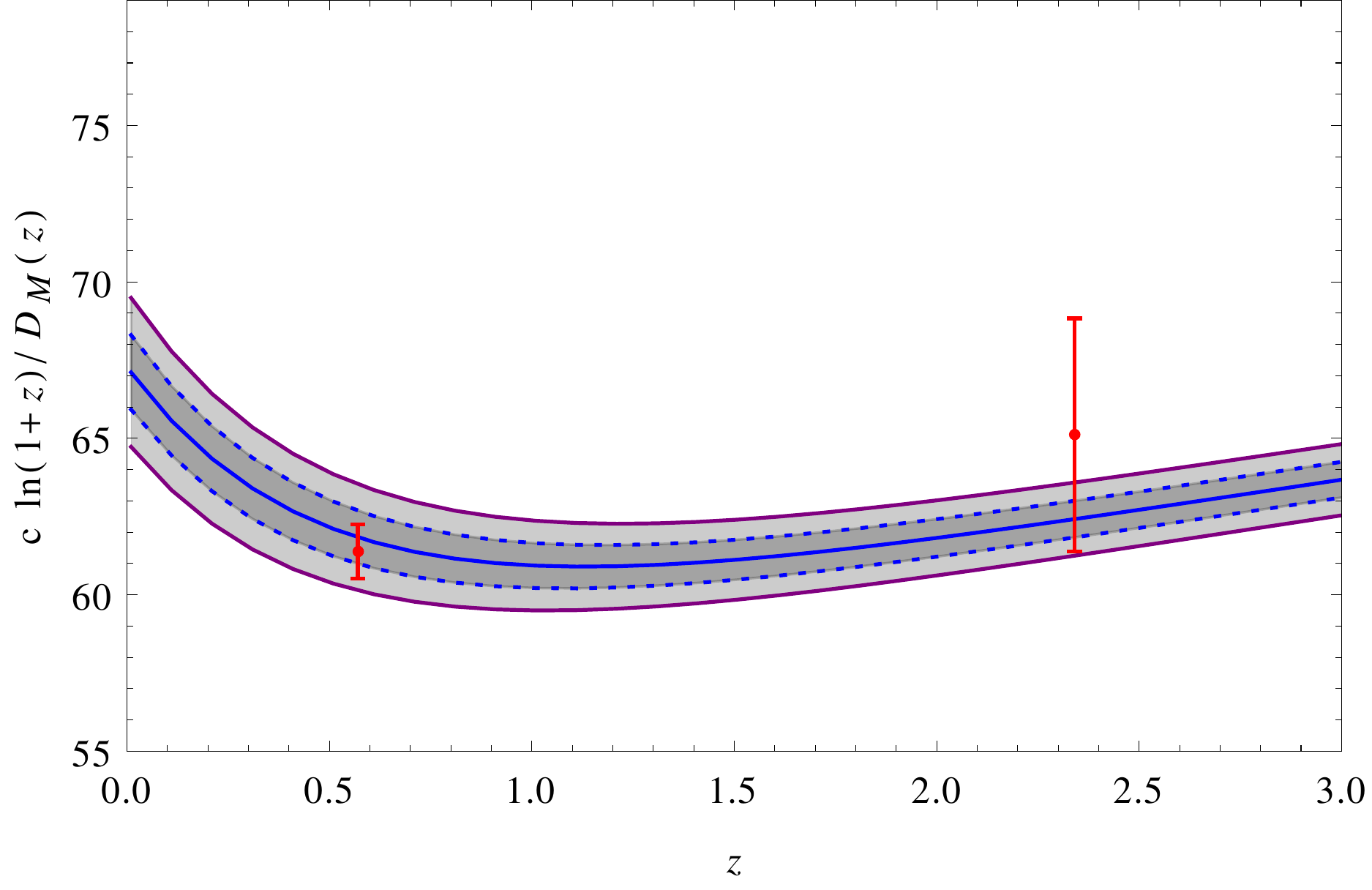}
\caption{\footnotesize{Plots of $c\ln(1+z)/D_M(z)$ curve  with 1$\sigma$ and 2$\sigma$ error bands for nonflat (left panel) and flat (right panel) $\Lambda$CDM models constrained by Planck+WP data. }}
\label{fig:4}
\end{figure}

\begin{figure}[htb!]\centering
\includegraphics[width=8.3cm]{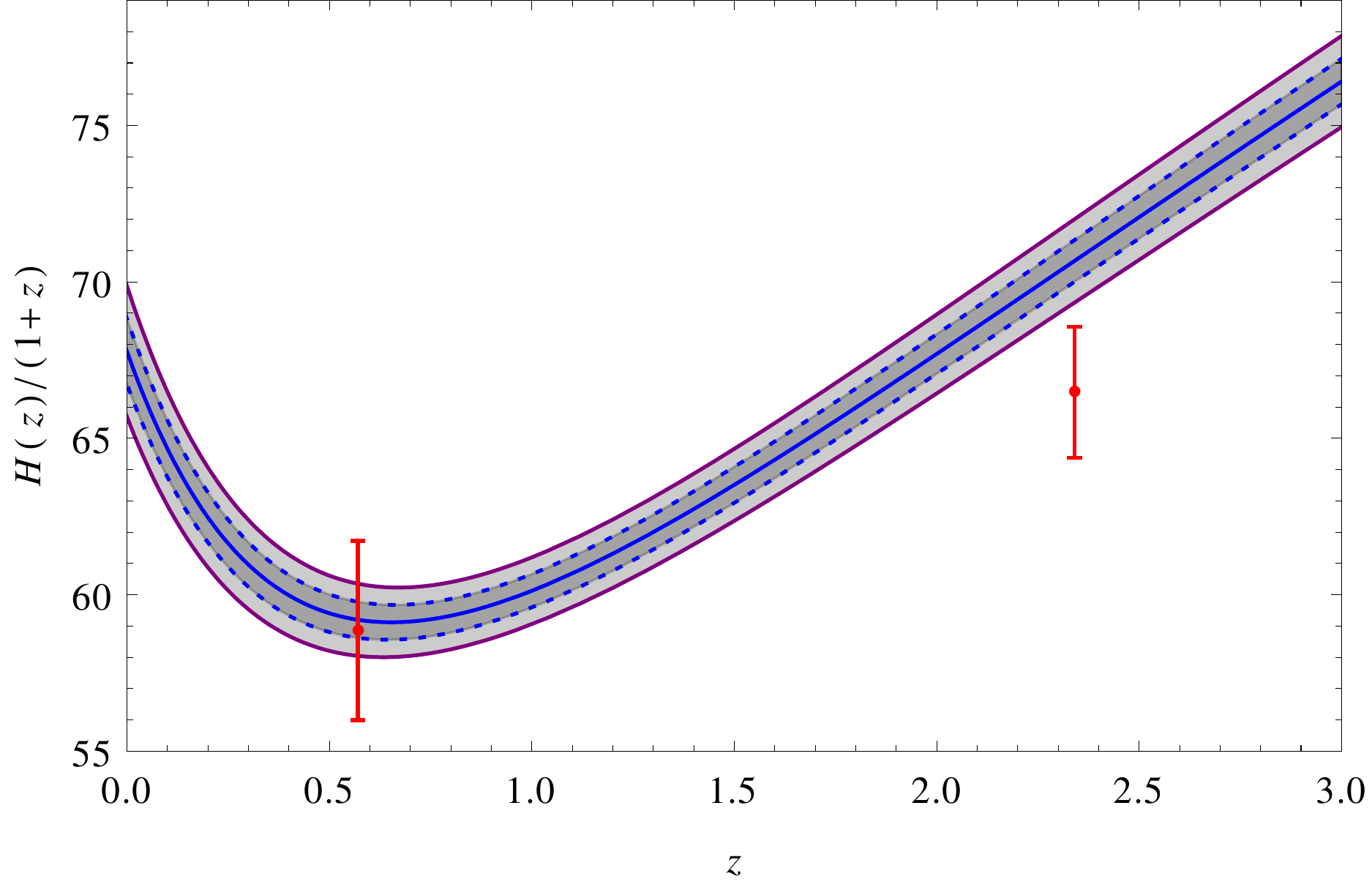}
\includegraphics[width=8.3cm]{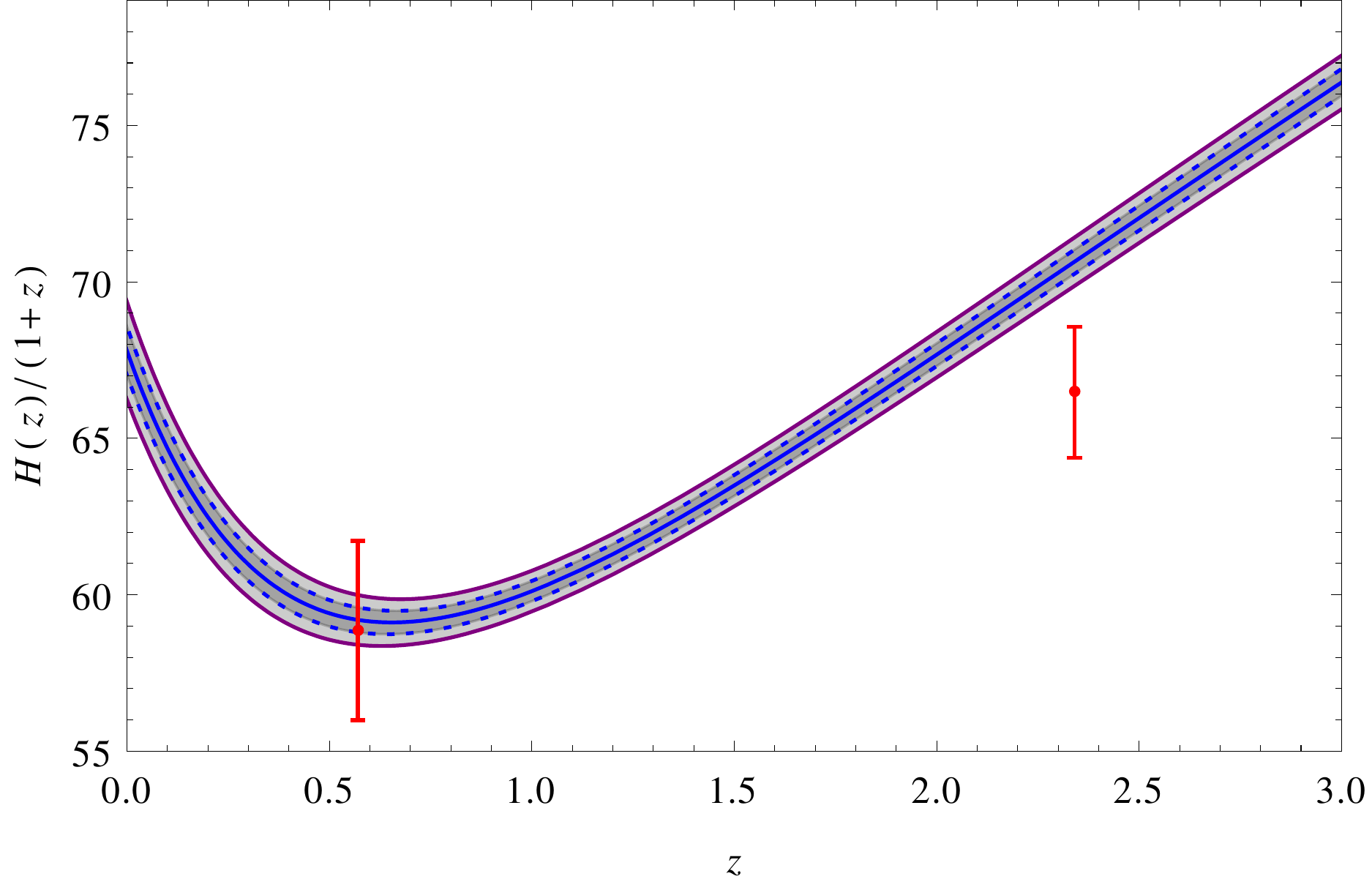}
\caption{\footnotesize{Plots of $H(z)/(1+z)$ curve with 1$\sigma$ and 2$\sigma$ error bands for nonflat (left panel) and flat (right panel) $\Lambda$CDM models constrained by Planck+WP+BAO data.}}
\label{fig:5}
\end{figure}

\begin{figure}[htb!]\centering
\includegraphics[width=8.3cm]{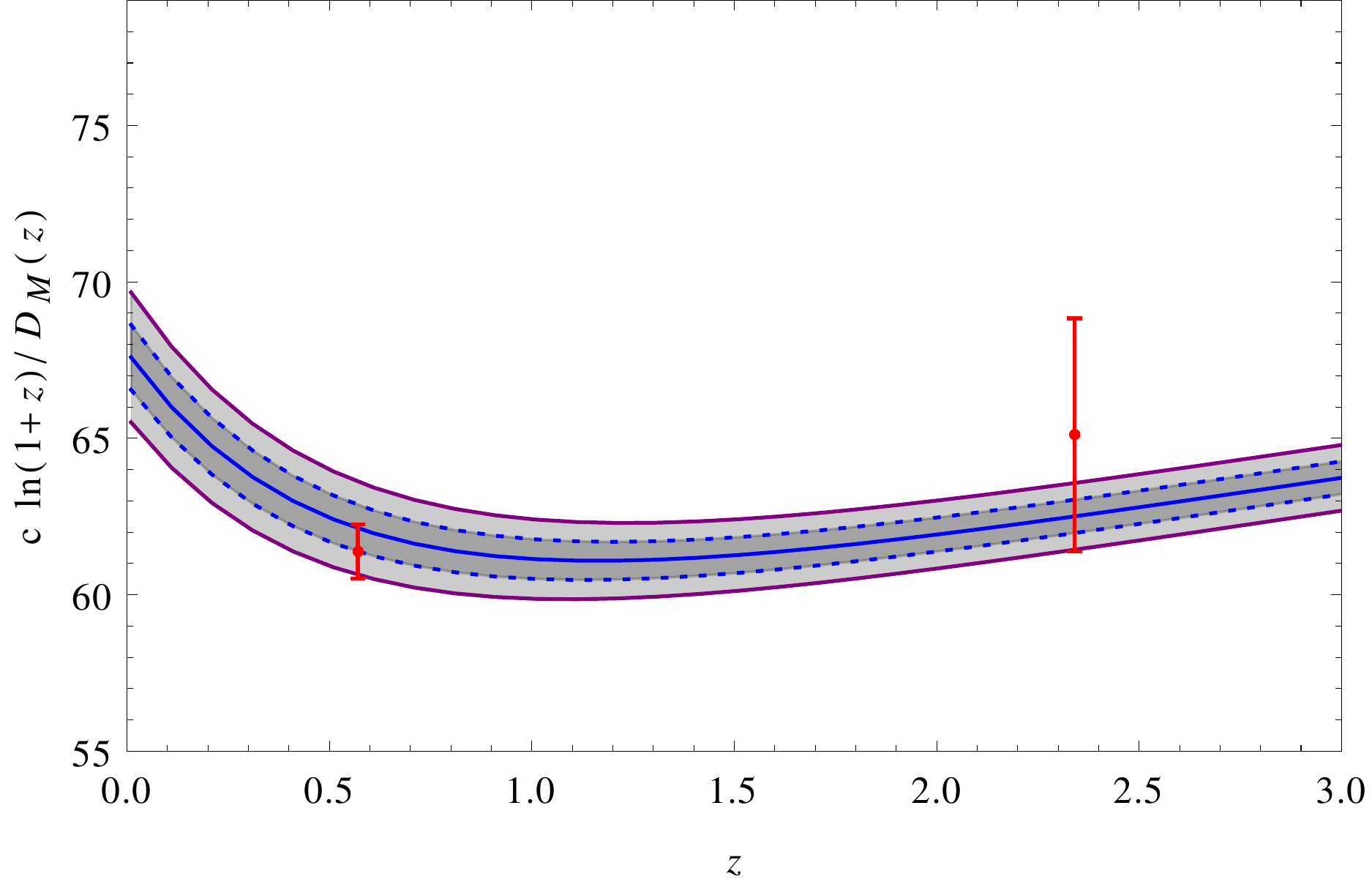}
\includegraphics[width=8.3cm]{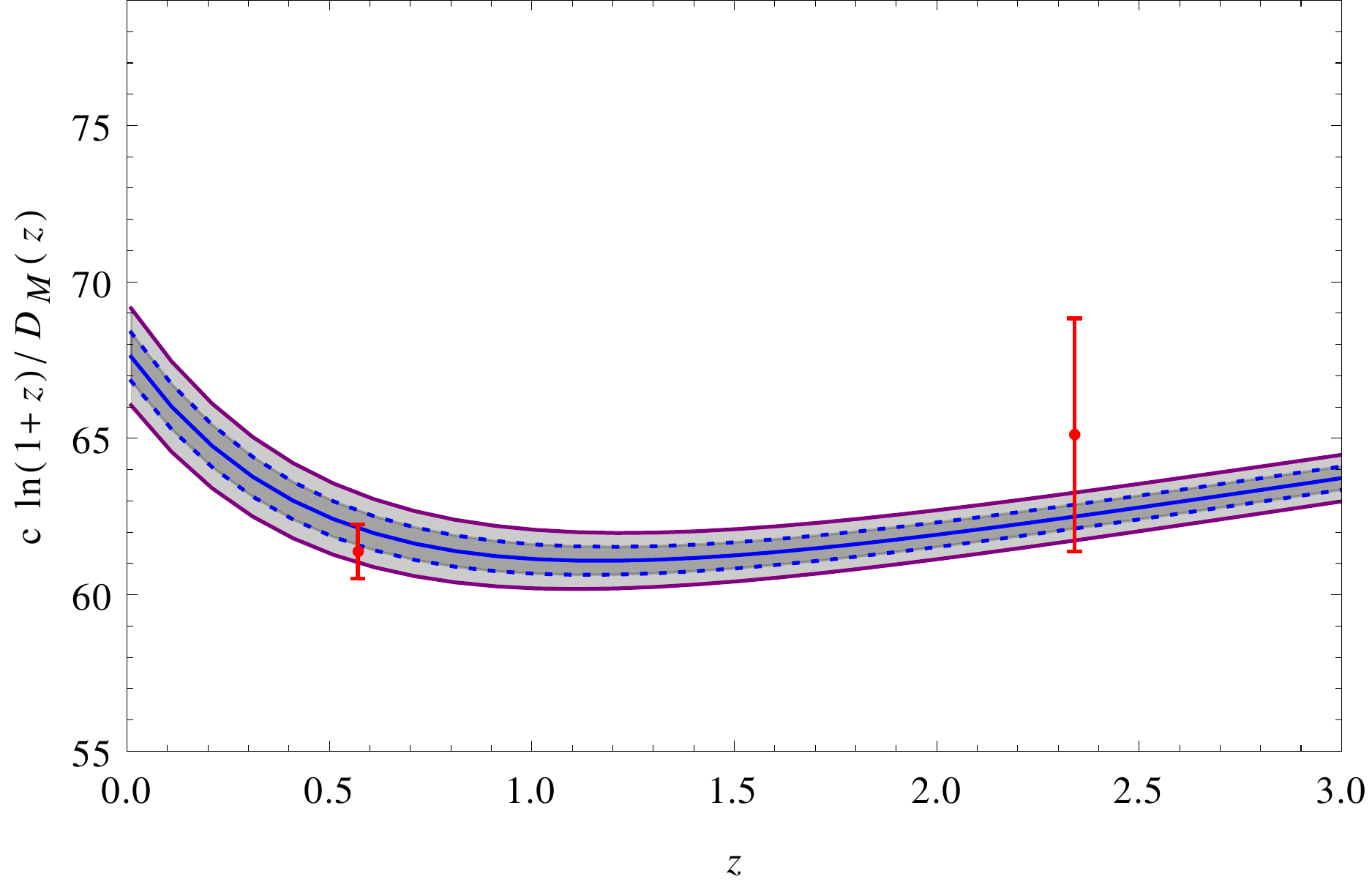}
\caption{\footnotesize{Plots of $c\ln(1+z)/D_M(z)$ curve  with 1$\sigma$ and 2$\sigma$ error bands for nonflat (left panel) and flat (right panel) $\Lambda$CDM models constrained by Planck+WP+BAO data. }}
\label{fig:6}
\end{figure}

We choose the BAO transverse measurements corresponding to redshifts $z=0.57$ (BOSS CMASS sample \cite{anderson14}) and $z=2.34$ (LyaF auto correlation \cite{delubac15}) as described in \cite{aub14}. In Figure \ref{fig:4}, we plot $c\ln(1+z)/D_M(z)$ curve  with 1$\sigma$ and 2$\sigma$ error bands for the nonflat (left panel) and flat (right panel) $\Lambda$CDM models constrained by Planck+WP data. From the plot in the left panel of Figure \ref{fig:4}, we observe that the nonflat $\Lambda$CDM model finds a poor fit to the BAO transverse measurements whereas the  right panel plot of Figure \ref{fig:4} shows that the flat $\Lambda$CDM models fits better to the transverse measurements. In order to realize the poor fit of nonflat $\Lambda$CDM model to the transverse BAO measurements, we further consider the constraints on nonflat $\Lambda$CDM model from Planck+WP+BAO data, and show the plots of $H(z)/(1+z)$ curve  with 1$\sigma$ and 2$\sigma$ error bands for the nonflat (left panel) and flat (right panel) $\Lambda$CDM models in Figure \ref{fig:5} while we show  the plots of $c\ln(1+z)/D_M(z)$ curve  with 1$\sigma$ and 2$\sigma$ error bands for the nonflat (left panel) and flat (right panel) $\Lambda$CDM models in Figure \ref{fig:6}. We notice that  the Planck+WP+BAO determined nonflat $\Lambda$CDM model finds poor fit to the high redshift  line of sight measurement while it exhibits good fit to the transverse measurements, as expected. Thus, the nonflat $\Lambda$CDM model constrained with Planck+WP data fits better to the line of sight measurement $H(z=2.34)=222\pm7$ km/s/Mpc, but only at the expense of still having a poor fit to the BAO transverse measurements.

The flat $\Lambda$CDM model with six free parameters is known to be the simplest best fit model to the presently available observational data. The nonflat $\Lambda$CDM model is simply an extension of the flat $\Lambda$CDM model with the additional free parameter $\Omega_k$. For the sake of comparison, we give the $\chi^2$ values in Table \ref{tab:results2} for the fitting of both the models to the two data set combinations used in our study. 
\begin{table}[h]
\caption{$\chi^2$ values for the flat and nonflat $\Lambda$CDM models.} 
\begin{center}
\begin{tabular}{|l|l|l|}
\hline Data&$\chi^2/2$ (Flat $\Lambda$CDM model)&$\chi^2/2$ (Nonflat $\Lambda$CDM model)\\ \hline 
Planck+WP	& 4903.927 & 4902.214\\
Planck+WP+BAO	& 4905.171 & 4904.691\\
\hline
\end{tabular}
\label{tab:results2}
\end{center}
\end{table}

We see that the nonflat $\Lambda$CDM model fits to both the data set combinations with almost the same $\chi^2$ values as the flat $\Lambda$CDM model does. It means one model can not be preferred over the other on the basis of $\chi^2$ values given in Table \ref{tab:results2}. However, if one considers the statistical analysis of the two models based on the information criteria such as Akaike Information Criterion (AIC), Kullback Information Criterion (KIC) and Bayes Information Criterion (BIC), then the nonflat $\Lambda$CDM model may be penalized because of the additional free parameter  $\Omega_k$ in comparison to the  flat $\Lambda$CDM model. For illustration of AIC, BIC, KIC and example models, one may refer to the Refs. \cite{melia13,ozgur14,kumar14}. Nevertheless, the nonflat $\Lambda$CDM model finds preference over the flat one when we look for the consistency of the two models with the line of sight BOSS measurement at higher redshift, as may be seen in Figure \ref{fig:3}. 

It is also not uncommon that a single, suitably chosen  data set, prefers (at $\sim 2 \sigma$ level) a model, which differs from the vanilla $\Lambda$CDM. However, when a global fit is performed, considering several different data sets, the non-standard model ends up not being preferred.  This is simply because the result for the single data sets may not really be statistically significant. So the observational data analysis carried out in this section in the context of nonflat $\Lambda$CDM model with the particular data set combination, namely Planck+WP, is just for the sake of illustration that the nonflat $\Lambda$CDM model constrained by this data set is capable of explaining the line of sight BOSS measurements at higher redshifts. It is also an indication that there are data sets which line up with the line of sight BOSS measurement when considered for consistency with the  nonflat $\Lambda$CDM model. Of course, there are data sets such as Planck+WP+BAO which give a negligible value of $\Omega_k$ and favor the flat $\Lambda$CDM model but are not able to explain the high redshift line of sight BOSS measurement as illustrated in this section. In short, the purpose of the present study is not to rule out the flat $\Lambda$CDM model rather to put forward the simple idea that the nonflat $\Lambda$CDM model with positive curvature could be a suitable model for explaining the high redshift line of sight BOSS measurement. Therefore, we point out that one should not necessarily resort to extreme theories of gravity with exotic dark energy (see the models described in Section \ref{2}) to explain a 2.5$\sigma$ deviation of the line of sight BOSS measurement from the flat $\Lambda$CDM model, when a simple curvature contribution can explain the deviation equally well.

\section{Concluding remarks}\label{5}
Following \cite{abdalla14}, we noticed that the lower values of $H(z)$ at higher redshifts can be achieved in phenomenological models of interacting dark sector as discussed in Subsection \ref{2.1}. Therefore, the new result from BOSS could be an indication of interaction in the dark sector as pointed out in \cite{abdalla14}. The lower values of $H(z)$ at higher redshifts can also be accommodated in modified $\Lambda$CDM model via screening mechanism as illustrated in Subsection \ref{2.2}, following \cite{sahni14}. Motivated by the screening idea, we have considered the nonflat $\Lambda$CDM model with positive curvature. We have found that this model successfully accommodates the lower value of $H(z)$ at higher $z$ suggested by BOSS. For the observational consistency check, we have considered the constraints on the nonflat $\Lambda$CDM model from the Planck+WP and Planck+WP+BAO data sets, which indicate that the nonflat $\Lambda$CDM model constrained with Planck+WP data fits better to the line of sight measurement $H(z=2.34)=222\pm7$ km/s/Mpc, but only at the expense of still having a poor fit to the BAO transverse measurements. Thus, in this work, we have briefly described some ideas/suggestions made in the literature to relieve the tension arising from the new result of BOSS, and showed that a nonflat $\Lambda$CDM model with positive curvature is a plausible choice to deal with the new line of sight measurement from BOSS in the sense that one does not necessarily need to resort to exotic dark energy models such as the ones described in Section \ref{2} to explain a 2.5$\sigma$ deviation of BOSS measurements from the flat $\Lambda$CDM model, when a simple curvature contribution can explain the deviation equally well.

As mentioned earlier, the nonflat $\Lambda$CDM model with considerable spatial curvature is in confrontation with the inflation paradigm which predicts almost a spatially flat Universe \cite{PlanckXXVI,aub14}. For instance, Linde \cite{linde03} pointed out that a realistic model of a spatially closed inflationary Universe is very difficult to produce even if the total number of e-folds is fine tuned to be sufficiently small. He proposed some artificial and fine-tuned models describing a spatially closed inflationary Universe, and concluded that though a broad class of inflationary models predict the spatial flatness of the Universe,  spatially closed Universes, if required, can be described without giving up all advantages of inflationary cosmology. He found that one can obtain  a semi-realistic model of a  spatially closed inflationary Universe with the fine-tuning at the level of about one percent, and such a fine-tuning would be much smaller and much easier in comparison to the one required for the explanation of
the enormously large mass, entropy, homogeneity and isotropy of the Universe, and also of the observed anisotropy of CMB without considering inflation. Thus, a fine-tuned bridging is possible between the inflationary scenario and the nonflat $\Lambda$CDM model with positive curvature.

In view of the consistency of the nonflat $\Lambda$CDM model with the line of sight high redshift measurement from BOSS as illustrated in Section \ref{3} and Section \ref{4}, one may interpret that it could be an indication of spatially nonflat Universe with positive curvature. It might be possible that we are estimating a value of cosmological constant screened by the curvature term. It is also  possible that the general relativity framework of $\Lambda$CDM model is not adequate, and requires modified theories of gravity to take into account \cite{sahni14}. In nutshell, if the nonflat $\Lambda$CDM model is true, then one needs to fine-tune the physics of inflation that can produce a Universe with spatial curvature. On the other hand, if the Universe is spatially flat, one needs to understand the physics of dark energy different from the cosmological constant. Considering overall scenario, it is clear that we do not have a model of the Universe in harmony with different observational probes. Nevertheless, the present study shows that inconsistency of the line of sight high redshift BOSS measurement with the flat $\Lambda$CDM model can be explained by simple curvature correction, and thereby it offers a plausible alternative to the proposals of exotic dark energy models suggested in the literature in order to resolve the said inconsistency. It is also pertinent to mention that the $\Lambda$CDM model with curvature correction finds a poor fit with the BAO transverse measurements whereas the exotic dark energy models suggested in the literature are yet to be tested for the transverse measurements.

\subsection*{Acknowledgments}
The author is grateful to the anonymous referee for valuable comments to improve the quality of the manuscript. The author is thankful to \"{O}zg\"{u}r Akarsu for fruitful discussions. The author is supported by the Department of Science and Technology, India under project No. SR/FTP/PS-102/2011. The author acknowledges the warm hospitality and research facilities provided by the Inter-University Centre for Astronomy and Astrophysics (IUCAA), India where a part of this work was carried out.


\begin{thebibliography}{99}
\bibitem{PlanckXXVI}
P. A. R. Ade et al., A \& A 571 (2014) A16 [arXiv:1303.5076 [astro-ph.CO]]
\bibitem{peebles03}
P. J. E. Peebles, B. Ratra, Rev. Mod. Phys. 75 (2003) 559 [arXiv:astro-ph/0207347]
\bibitem{copeland06}
E. J. Copeland, M. Sami, S. Tsujikawa, Int. J. Mod. Phys. D 15 (2006) 1753 [arXiv:hep-th/0603057]
\bibitem{verde14} 
L. Verde, P. Protopapas, R. Jimenez, Phys. Dark Univ. 5-6 (2014) 307. [arXiv:1403.2181 [astro-ph.CO]]
\bibitem{delubac15} 
T. Delubac et al., A \& A 574 (2015) A59 [arXiv:1404.1801 [astro-ph.CO]]
\bibitem{samushia13} 
L. Samushia et al., MNRAS 429 (2013) 1514 [arXiv:1206.5309 [astro-ph.CO]]

\bibitem{he09} 
J. -H. He, B. Wang, E. Abdalla, Phys. Lett. B 671 (2009)
139  [arXiv:0807.3471 [gr-qc]]
\bibitem{costa14}
A. A. Costa, X. D. Xu, B. Wang, E. G. M. Ferreira, E. Abdalla, Phys. Rev. D 89 (2014) 103531 
[arXiv:1311.7380 [astro-ph.CO]]
\bibitem{abdalla14} 
E. Abdalla et al., arXiv:1412.2777 [astro-ph.CO]

\bibitem{sahni14} 
V. Sahni, A. Shafieloo, A. A. Starobinsky, ApJ 793 (2014) L40 [arXiv:1406.2209 [astro-ph.CO]].
\bibitem{sahni03} 
V. Sahni, Yu. Shtanov, JCAP 0311 (2003) 014 [arXiv:astro-ph/0202346]
\bibitem{aub14} 
E. Aubourg et al., arXiv:1411.1074 [astro-ph.CO]
\bibitem{bond97}
J. R. Bond, G. Efstathiou, M. Tegmark, MNRAS 291 (1997) L33
[arXiv:astro-ph/9702100]

\bibitem{zal97}
M. Zaldarriaga, U. Seljak, Phys. Rev. D 55 (1997) 1830 [arXiv:astro-
ph/9609170]
\bibitem{stom99}
R. Stompor, G. Efstathiou,  MNRAS 302 (1999) 735 [arXiv:astro-ph/9805294]
\bibitem{sper07}
D. N. Spergel et al., ApJS 170 (2007) 377 [arXiv:astro-ph/0603449]
\bibitem{Sievers13} 
J. L. Sievers et al., JCAP 10 (2013) 060 [arXiv:1301.0824 [astro-ph.CO]]


\bibitem{van12}
A. van Engelen et al. ApJ 756 (2012) 142 [arXiv:1202.0546]
\bibitem{Planck15}
P. A. R. Ade et al., arXiv:1502.01589 [astro-ph.CO]
\bibitem{cos02}
http://cosmologist.info/cosmomc/; A. Lewis and S. Bridle, Phys. Rev. D 66, 103511 (2002) [arXiv:astro-
ph/0205436]

\bibitem{anderson14}
L. Anderson et al.,  MNRAS 441 (2014) 24 [arXiv:1312.4877 [astro-ph.CO]]
\bibitem{melia13}
F. Melia and R. S. Maier, MNRAS 432 (2013) 2669 [arXiv:1304.1802 [astro-ph.CO]]
\bibitem{ozgur14} 
\"O. Akarsu, S. Kumar, R. Myrzakulov, M. Sami, L. Xu, JCAP 01 (2014) 022 [arXiv:1307.4911 [gr-qc]]
\bibitem{kumar14} 
S. Kumar, Mod. Phys. Lett. A 29 (2014) 1450119 [arXiv:arXiv:1404.1910 [gr-qc]]


\bibitem{linde03}
A. Linde, JCAP 0305 (2003) 002 [arXiv:astro-ph/0303245]

\end{thebibliography}
\end{document}